\def\year{2023}\relax
\documentclass[letterpaper]{article} 
\usepackage{aaai22}  
\usepackage{times}  
\usepackage{helvet}  
\usepackage{courier}  
\usepackage[hyphens]{url}  
\usepackage{graphicx} 
\usepackage{titlecaps}
\usepackage{subfigure}
\usepackage{booktabs}
\usepackage{amsmath}
\usepackage{natbib}  
\usepackage{caption} 
\DeclareCaptionStyle{ruled}{labelfont=normalfont,labelsep=colon,strut=off} 
\usepackage{algorithm}
\usepackage{algorithmic}
\usepackage{threeparttable}

\usepackage{newfloat}
\usepackage{listings}
\floatstyle{ruled}
\newfloat{listing}{tb}{lst}{}
\floatname{listing}{Listing}
\title{Sequential Graph Attention Learning for Predicting Dynamic Stock Trends \\ (Student Abstract)}
\author{
     Tzu-Ya Lai\textsuperscript{\rm 1},
    Wen Jung Cheng\textsuperscript{\rm 2},
     Jun-En Ding\textsuperscript{\rm 3}\thanks{Corresponding Author\\}}
\affiliations{
    \textsuperscript{\rm 1} National Taipei University Master of Arts in Economics\\

     \textsuperscript{\rm 2} University of Connecticut Master in Financial Technology\\
    \textsuperscript{\rm 3} National Yang Ming Chiao Tung University Institute of Hospital and Health Care Administration

    s710961103@gm.ntpu.edu.tw, wen-jung.cheng@uconn.edu, m02040013@nycu.edu.tw \\
    
}

\usepackage{bibentry}

\begin{document}
\maketitle
\begin{abstract}
The stock market is characterized by a complex relationship between companies and the market. This study combines a sequential graph structure with attention mechanisms to learn global and local information within temporal time. Specifically, our proposed “GAT-AGNN” module compares model performance across multiple industries as well as within single industries. The results show that the proposed framework outperforms the state-of-the-art methods in predicting stock trends across multiple industries on Taiwan Stock datasets.
\end{abstract}

\section{Introduction}
The stock market can be considered a dynamic system with numerous intricately connected parts that change over time. Graph Attention Networks (GATs) and Gated Recurrent Units (GRUs) have been used in some research to capture temporal information of the stock exchange due to the rapid growth of Graph Neural Networks (GNNs) (\citeauthor*{hsu2021fingat}, \citeyear{hsu2021fingat}). Each company is considered a node in the graph in these studies, and the correlation between two nodes at a certain time point determines the edge between them. In order to aggregate data on momentum spillover between companies, \citeauthor*{cheng2021modeling} (\citeyear{cheng2021modeling}) previously proposed a model that integrates Graph Attention Network (GAT) and Attribute-Mattered Aggregator. In this paper, we argue that: a) the majority of research does not confirm the efficacy of the graph neural network approach, including whether it can be data-driven to capture company interrelationships; and b) the single-layer feed-forward neural network used as the Attribute-Mattered Aggregator in previous research (\citeauthor*{cheng2021modeling}, \citeyear{cheng2021modeling}) should be converted to a more methodologically sophisticated neural network module. We propose a novel model framework called the GAT-AGNN model to enhance prediction performance. We also suggest implementing the Corr-Cos model to confirm the efficacy of the graph neural network approach.

\subsubsection{Problem Definition}
This study aims to forecast return on stocks across multiple industries based on their histories. We focus on two tasks founded in our analysis of return ratios from daily trading information: a) classification problems (classifying the future stock return from the closing price as a positive return or negative return); and b) regression problems (predicting future stock closing price return).

\subsubsection{Data Collection}
Our research uses daily data from \emph{Taiwan Economic Journal} (TEJ) to predict the stock market trend. Every company has a highest/lowest price, opening/closing price and turnover rate. Data was collected from November 18, 2019, to June 20, 2022; a total of 630 days. The 30 companies studied were selected from the Yuanta/P-shares Taiwan Mid-Cap 100 ETF.

\section{Proposed Method}
\subsection{Sequential Fusion Module with GAT and AGNN}
\subsubsection{Gated Recurrent Unit (GRU)}
Given the $i-th$ company, and the independent variable for time series representation learning is $\mathcal{X}_{i}^{[t-T:t)}=[\begin{bmatrix}\mathbf{x}_{i}^{t-T},&\cdots,&\mathbf{x}_{i}^{t-1}\end{bmatrix}]^{\mathsf{T}}$ which is the historical data for the first $T$ days of the $i-th$ firm, where $T$ is the length of the rolling window. This research adopts GRU as the module for time series data extraction.

\begin{equation}\label{eq:1} 
    \mathbf{h}_{i}^{t}= GRU(\mathcal{X}_{i}^{[t-T:t)}).
\end{equation}

The historical feature vector $\mathbf{h}_{i}^{t}$ will be used to retrieve data for the previous $T$ days in accordance with the approach described above. 

Our approach utilizes the GRU module to create a graph-structured dynamic system (see Figure 1) that captures time-series dependency information for each company, and the nodes that constitute this dynamic system are derived from the fusion historical information vectors $\mathcal{H}=[\begin{bmatrix}{\mathbf{h}_{1}^{t}},&\cdots,&{\mathbf{h}_{N}^{t}} \end{bmatrix}]$.

\subsubsection{Graph Attention Network (GAT)}
The prospective link between companies i and j at time t is therefore captured by ${U_{i,j}}^{t}$.
\begin{equation} \label{eq:3} 
	\mathcal{U}_{i,j}^{t} = u(\mathbf{h}_{i}^{t},\mathbf{h}_{j}^{t})
	                      = ELU({\mathbf{a}_{u}}^{\mathsf{T}} \mathcal{W}_{u}
	                          [{\mathbf{h}_{i}^{t}}^{\mathsf{T}} \mathbin\Vert {\mathbf{h}_{j}^{t}}^{\mathsf{T}}] ),
\end{equation}
To identify prospective company relationships via the shared attention mechanism, a single-layer neural network and an ELU activation function are used to train the correlation between nodes  $i$ and $j$. The softmax function is applied to normalize all pairwise connections, and the correlation between companies  $(i,j)$, is normalized at time $t$, can be identified as follows:

\begin{equation} \label{eq:4} 
	\widetilde{Q_{i,j}}^{t}=softmax_{j}= \frac{exp(\mathcal{U}_{i,j}^{t})}
	                               { \sum_{k{\in}N ,k{\neq}i} exp(\mathcal{U}_{i,k}^{t})}.
\end{equation}

\begin{figure}[H]
\centering  
\subfigure{
\label{Fig.1}
\includegraphics[width=0.46 \textwidth, height=0.2\textwidth]{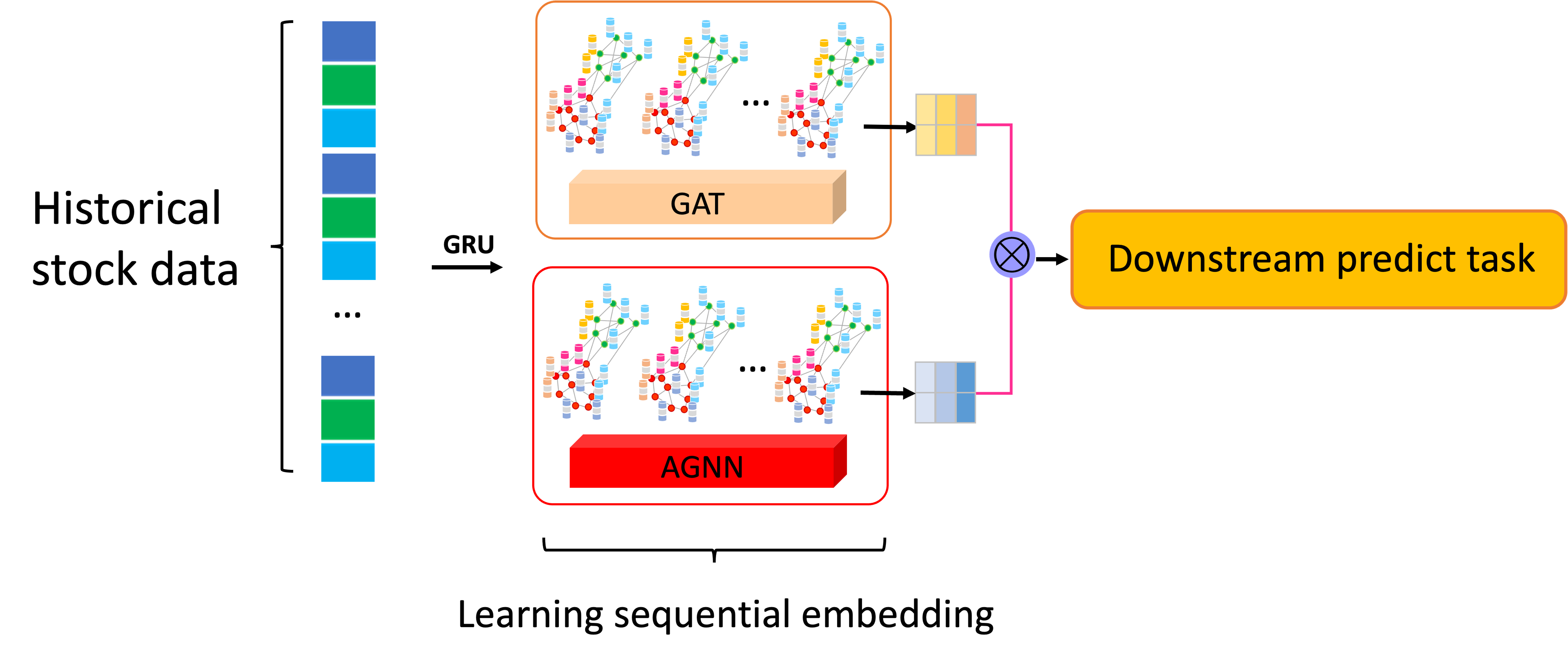}}
\caption{Diagram of GAT-AGNN’s architecture.}
\end{figure}

\subsubsection{Attention-Based Graph Neural Network (AGNN)}

Using calculation ${C_{i,j}}^{t}$ can capture potentially connected organizations based on similarities between companies. By using a trainable $\alpha$, the prospective connections between companies $i$ and $j$ at time $t$, can be identified as follows:

\begin{equation} \label{eq:5} 
	\mathcal{C}_{i,j}^{t} = c(\mathbf{h}_{i}^{t},\mathbf{h}_{j}^{t})
	                      = \alpha \cos( \mathcal{W}_{c}
	                          [{\mathbf{h}_{i}^{t}}^{\mathsf{T}}, {\mathbf{h}_{j}^{t}}^{\mathsf{T}}] ),
\end{equation}

The softmax function is used to normalize all pairwise connections.

\begin{equation} \label{eq:6} 
	\widetilde{G_{i,j}}^{t}=softmax_{j}= \frac{exp(\mathcal{C}_{i,j}^{t})}
	                               { \sum_{k{\in}N ,k{\neq}i} exp(\mathcal{C}_{i,k}^{t})}.
\end{equation}

\subsubsection{Sequential Relation Module}
The relational of the $i-th$ company at time $t$ information from
the GRU module vector is defined as $\mathcal{V}=[\begin{bmatrix}{\mathbf{v}_{1}^{t}},&\cdots,&{\mathbf{v}_{N}^{t}} \end{bmatrix}]$, while  ${\mathbf{v}_{i}^{t}}$ is the correlation sequential vector calculated as follows:

\begin{equation} \label{eq:2} 
    \mathbf{v}_{i}^{t}={ \boldsymbol{tanh}( \sum_{j,j{\neq}i}^N  \widetilde{Q_{i,j}}^{t}  W_{s}  
    {\mathbf{h}_{j}^{t}}^{\mathsf{T}}\otimes
\widetilde{G_{i,j}}^{t}  W_{v}  
    {\mathbf{h}_{j}^{t}}^{\mathsf{T}})},
\end{equation}

In addition to the attention mechanism, we consider a multi-head option.
\begin{equation} \label{eq:7} 
 \mathbf{v}_{i}^{t}=\mathop{\Big\Vert}\limits_{m=1}^{M}
 { \boldsymbol{tanh}( \sum_{j,j{\neq}i}^N  
{\widetilde{Q_{i,j}}}^{t,[m]}{W_{s}}^{[m]}   {\mathbf{h}_{j}^{t}}^{\mathsf{T}}
\otimes
{\widetilde{G_{i,j}}}^{t,[m]}{W_{v}}^{[m]}  {\mathbf{h}_{j}^{t}}^{\mathsf{T}})},
\end{equation}

The purpose of our prediction tasks are both Regression and Classification tasks. First, a single-layer neural network with the function $RELU({\mathcal{W}_{i}}^{\mathsf{T}}
                [{\mathbf{h}_{i}^{t}}^{\mathsf{T}} \mathbin\Vert {\mathbf{v}_{i}^{t}}]
                +\mathbf{b}_{i})$ is used; then, the second uses a single-layer neural network with the function $Softmax({\mathcal{W}_{i}}^{\mathsf{T}}
                [{\mathbf{h}_{i}^{t}}^{\mathsf{T}} \mathbin\Vert {\mathbf{v}_{i}^{t}}]
                +\mathbf{b}_{i})$.

Additionally, we replace the Sequential Relation Module
of GAT and AGNN with the without training parameters module of Correlation coefficient and Cosine similarity (Corr-Cos model). With these settings, we can test which modules are more robust in their ability to learn complex information from industry relationships.

\begin{equation} \label{eq:16} \nonumber
    \mathbf{v}_{i}^{t}={ \boldsymbol{tanh}( \sum_{j,j{\neq}i}^N  \widetilde{Corr_{i,j}}^{t}  W_{s}  
    {\mathbf{h}_{j}^{t}}^{\mathsf{T}}\otimes
\widetilde{Cos_{i,j}}^{t}  W_{v}  
    {\mathbf{h}_{j}^{t}}^{\mathsf{T}})}.
\end{equation} 

\begin{table}[htb!]
\centering
\setlength{\tabcolsep}{1mm}{
\begin{tabular}{l|ll|ll}
\hline
Problems&Classification&&Regression\\
\cline{2-5}
Metrics&ACC&AUC&MSE&MAE\\
\hline
\hline
LSTM&0.5222&0.5154&1.7299&0.9591\\
LSTM+GAT&0.5283&0.5023&1.3874&0.8633\\
GRU&0.5139&0.5027 &1.4963& 0.8944\\
GRU+GAT&0.5150&0.4805&1.2083& 0.8031\\
Corr-Cos&0.5328&\bf{0.5290}&1.1850&0.7940\\
AD-GAT& 0.5189& 0.5056&1.2143&0.8183\\
GAT-AGNN&\bf{0.5350}&0.5190&\bf{1.1585}& \bf{0.7823}\\
\hline
\end{tabular}}
\caption{Comparison of classification and regression prediction results across models.}
\label{table1}
\end{table}

\section{Results and Discussion}
In our research, we utilize different creating relationship functions (2) and (4) to enhance and combine the attention mechanisms of GAT and AGNN for the multi-head setting, which is superior to the traditional approach that relies only on an activate function with a single-layer neural network.

\begin{table}[htb!]
\centering
\begin{threeparttable}
\begin{tabular}{lll}
\hline
 &ACC&MSE\\
\hline
All Industry Industry6 \tnote{1}&\bf{0.5700}&\bf{0.9851}\\
\midrule
Industry6\tnote{1}&0.4767&1.2542\\
\hline
\end{tabular}  
\begin{tablenotes}
        \footnotesize
        \item[1] Industry 6: shipping industry.
       
\end{tablenotes}
\end{threeparttable}
\caption{Comparison of model performance across multiple industries and within single industries.}
\end{table}

\begin{figure}[H]
\centering  
\subfigure{
\label{Fig.2}
\includegraphics[width=0.36\textwidth]{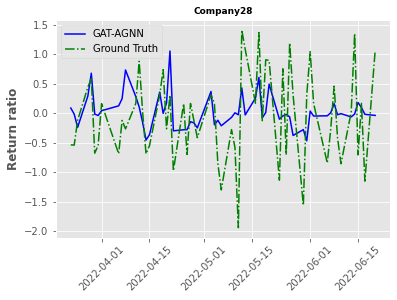}}
\caption{Stock time series forecasting visualization. Time-dependent pattern captured by GAT-AGNN.}
\end{figure}

The results on the classification and regression prediction tasks are displayed in Table1. According to the results below, the GAT-AGNN model outperforms the Corr-Cos model, indicating that graph neural networks are effective for such tasks. Moreover, the GAT-AGNN model lowers MSE by $5.5\%$ over the AD-GAT model (\citeauthor*{cheng2021modeling}, \citeyear{cheng2021modeling}). Table1 shows, the GAT-AGNN model is not only significantly superior to GRU and GRU+GAT but also LSTM and LSTM+GAT.

It is evident in Table 2 that the MSE in the multiple-industry model decreases by 26.91\%, while the ACC increases by 9.33\% compared with the single-industry model. Thus, our study suggests that sequential graph structure in hierarchical information (see Figure 2) offers promise for the identification of industry relations and information.
\section{References}
\nobibliography*
\bibentry{cheng2021modeling}.\\[.2em]
\bibentry{hsu2021fingat}.\\[.2em]

\nobibliography{aaai22}
\end{document}